\documentclass{cimento}

\usepackage{graphicx} 
\usepackage{amssymb,amsmath,amsthm,amsfonts}
\usepackage{braket}
\usepackage[compat=1.1.0]{tikz-feynman}
\usepackage{comment}

\title{Low-energy flavour probes of light vector bosons}
\author{L.~Di Luzio \from{ins:x}, G.~Levati \from{ins:y}\from{ins:x}\thanks{Speaker}, P.~Paradisi \from{ins:y}\from{ins:x}
        \atque
X.~Ponce D\'iaz \from{ins:y}\from{ins:x}}

\instlist{\inst{ins:x} INFN, Sezione di Padova - Via Marzolo 8, 35131, Padova, Italy
\inst{ins:y} Dipartimento di Fisica e Astronomia `G.~Galilei', Universit\`a di Padova  - Via Marzolo 8, 35131, Padova, Italy, }

\PACSes{\PACSit{12.39.Fe}{Chiral Lagrangians}
\PACSit{13.25.-k}{Hadronic decays of mesons}
\PACSit{12.60.-i}{Models beyond the standard models}}

\begin{document}

\maketitle

\begin{abstract}
In this work, we construct the chiral Lagrangian for a light spin-1 boson $X$ possessing both vectorial and axial couplings to the light Standard Model quarks $u, d, s$.
We then use it in order to describe the tree-level, model-independent contributions to the $\Delta S = 1$ transition $K^\pm \rightarrow \pi^\pm  X$, which is induced by Standard Model charged currents and is possibly enhanced by the emission of a longitudinally polarized $X$ boson.
Such a flavour observable is then shown to set the best model-independent bounds on the diagonal axial couplings of $X$ to light quarks in the mass range allowed by the decay kinematics, improving the currently available constraints from beam-dump experiments and collider searches.
\end{abstract}

\section{Introduction}

The lack of any detection of heavy New Physics (NP) at the LHC has been pushing the theoretical community to explore new Beyond the Standard Model (BSM) physics scenarios. 
These generically consider either new particles that are too heavy to be possibly detected at collider experiments, or focus on new light and feebly interacting massive particles that have so far gone undetected.
The new BSM particles introduced in the second scenario have been receiving a steadily  increasing attention, both from a theoretical point of view and from an experimental one.

Several studies in this direction were devoted to the analysis of the properties of a hypothetical ``dark photon'', a new massive spin-1 boson which is kinetically mixed with the ordinary photon and whose interactions with SM particles can act as a portal to a dark sector ~\cite{Holdom:1985ag,Pospelov:2007mp}.
Efforts in detecting the dark photon include beam-dump~\cite{Bjorken:2009mm}, 
fixed-target~\cite{APEX:2011dww,Merkel:2014avp}, 
collider~\cite{BaBar:2009lbr,Curtin:2013fra,BaBar:2014zli,LHCb:2017trq,Anastasi:2015qla,LHCb:2019vmc,CMS:2019buh}, and meson decay~\cite{Bernardi:1985ny,KLOE-2:2011hhj,KLOE-2:2012lii,WASA-at-COSY:2013zom,HADES:2013nab,NA482:2015wmo,NA62:2019meo} experiments.

Generalisations of the dark photon scenario featuring a light spin-1 boson $X$ possessing general couplings to SM fermions have been envisaged and analysed as well (see e.g.~\cite{Kahn:2016vjr,Ilten:2018crw,Baruch:2022esd,Asai:2022zxw}). Interestingly enough, if the $X$ couples to non-conserved currents of SM fields, processes involving its longitudinal component will result in being possibly enhanced by the ratio $($energy$/m_X)^2$ \cite{Dror:2017ehi,Dror:2017nsg}, thus amounting for the largest contribution to the related observables.

In this article we will discuss, based on our work in \cite{DiLuzio:2023xyi}, the sensitivity of this scenario to the rare flavour-changing process $K^\pm \rightarrow \pi^\pm X$.
In order to do so, we will show how to build the most general $\Delta S = 1$ Chiral Lagrangian up to order $\mathcal{O} (p^4)$ that is necessary to account for all of the weak-induced flavour transitions $s \rightarrow d$ prompting the aforementioned decay process.

The weak-induced flavour-changing interactions one has to consider fall in either one of two categories: they can be either $\mathcal{O} (p^2)$ terms stemming from an effective $sdX$ vertex  generated by the one-loop exchange of a W boson and up-type quarks ~\cite{Dror:2017ehi,Dror:2017nsg}, or they can be $\mathcal{O} (p^4)$ contributions arising from the tree-level initial- or final-state radiation of an $X$ boson from external quark legs.
We will show that the two contributions are comparable in size, the former being of lower order in the Chiral expansion but necessarily arising at one-loop level, while the latter is a tree-level one appearing however only at next-to-leading order in the Chiral expansion, \textit{i.e.} when four-fermion $\Delta S  = 1$ operators are included in the Lagrangian.
The tree-level, $\mathcal{O} (p^4)$ contributions moreover have the virtue of being model-independent, therefore representing a robust prediction of any Ultraviolet (UV) complete NP model predicting the existence of extra $U(1)$ light spin-1 bosons. The loop-induced effects discussed in ~\cite{Dror:2017ehi,Dror:2017nsg} are instead sensitive to the specific realisation of the UV completion mechanism providing the $X$ boson with a mass.

\section{$\Delta S = 1$ chiral Lagrangian for spin-1 bosons}

The most general Lagrangian describing the interactions of a new spin-1 boson $X$ with the SM light quarks $q = (u,d,s)^T$ can be written as
\begin{equation}
\label{eq:X_Lag_with_SM_Fermions}
\mathcal{L}^{\rm int}_X = g_x X_\mu \, \bar{q}\,\gamma^\mu (x_V + x_A \gamma_5) \, q \, , 
\end{equation}

where $g_x$ measures the strength of the universal coupling of $X$ to quarks. The vectorial and axial charges, $x_{V,A}$, are matrices in flavour space and may include off-diagonal entries in the 2-3 sector.

\subsection{Lowest-order chiral Lagrangian}

The description we have outlined in the previous section is of course valid at energies above few GeV, where the Lagrangian in eq. \eqref{eq:X_Lag_with_SM_Fermions} can be directly employed to analyse the interactions of $X$ with quarks. 

Below the QCD scale however quarks confine and they are no longer the most adequate degrees of freedom for describing physical processes and one should rather resort to a description in terms of mesons and baryons. 
In order to discuss the interactions of mesons and baryons with other particles one can then make use of Chiral Perturbation theory ($\chi$PT) techniques~\cite{Gasser:1984gg,Pich:1995bw}.

In particular, the interaction of an extra spin-1 boson $X$ with quarks can be implemented in a $\chi$PT setup as follows:
first one considers the massless QCD Lagrangian with chiral symmetry group $G = SU(3)_L \times SU(3)_R$ 
\begin{equation}
\label{eq:Free_QCD_Lag}
\mathcal{L}_{\text{QCD}}^0 = 
-\frac{1}{4} G^a_{\mu\nu}G_a^{\mu\nu} 
+ i \bar{q}_L \gamma^\mu 
\left(\partial_\mu + ig_s\frac{\lambda_a}{2}A^a_\mu\right) q_L
+ i \bar{q}_R \gamma^\mu 
\left(\partial_\mu + ig_s\frac{\lambda_a}{2}A^a_\mu \right) q_R \, ,
\end{equation}  
where $q = (u, d, s)^T$ and $\lambda_a$ are the Gell-Mann matrices.

Chiral symmetry-breaking terms (like mass terms or interactions with external gauge fields other than gluons) can be implemented 
by introducing appropriate spurions ($r_\mu,\, l_\mu,\,$ $ s, \,p$) as external source fields \cite{Gasser:1984gg}. 
The resulting Lagrangian  
$\mathcal{L}_{\text{QCD}}^{\text{ext}}$ then reads
\begin{equation}
\label{eq:QCD_Lag_With_Sources}
\begin{split}
\mathcal{L}_{\text{QCD}}^{\text{ext}} 
&= \mathcal{L}_{\text{QCD}}^0 + \bar{q} \gamma^\mu (2 r_\mu P_R + 2 \ell_\mu P_L)q + \bar{q} (s - i p \gamma_5) q\,.
\end{split}
\end{equation}

Its chiral counterpart is found to be 
\begin{equation}
\label{eq:Chipt_Lag_With_Sources}
\mathcal{L}_{\chi \text{PT}}^{\text{ext}} = \frac{f_\pi^2}{4} \,\text{Tr}\left[D_\mu U^\dagger D^\mu U + U^\dagger \chi + \chi^\dagger U\right] +\mathcal{O} (p^4)
\end{equation}
where $U (x) = \exp \left[ i \lambda_a \pi_a(x)/f_\pi \right]$ 
is the mesonic matrix transforming as $U(x) \rightarrow L U(x) R^\dagger$ under $SU(3)_L \times SU(3)_R$ and $\pi_a(x)$ are the Goldstone boson fields of $SU(3)_L \times SU(3)_R \to SU(3)_{V}$ spontaneous breaking. Moreover, the following quantities have been defined:
\begin{equation}
 D_\mu U = \partial_\mu U - i r_\mu U + i U \ell_\mu  \qquad \text{and} \qquad \chi = 2B_0 \, (s + i p)\,.
\end{equation}

In the model described by eq.\eqref{eq:X_Lag_with_SM_Fermions}, the covariant derivative $D_\mu U$ reads
\begin{equation}
    D_\mu U= \partial_\mu U -i g_x X_\mu(Q^x_R U-U Q^x_L) \, , 
\end{equation}
where $Q^x_{R/L}=Q^x_V\pm Q^x_A$, while 
\begin{equation}
Q^x_V = \begin{bmatrix} x_V^u & 0 & 0 \\ 0 & x_V^d & x_V^{23}\\ 0& x_V^{32} & x_V^s\end{bmatrix} \qquad \text{and} \qquad Q^x_A = \begin{bmatrix} x_A^u & 0 & 0 \\ 0 & x_A^d & x_A^{23}\\ 0& x_A^{32} & x_A^s\end{bmatrix} 
\end{equation}
The Lagrangian in \eqref{eq:Chipt_Lag_With_Sources} can then be expanded in terms of the constituent meson fields. The lowest order terms in the NP coupling relevant to $K^\pm \rightarrow \pi^\pm X$  read
%
\begin{align}
\label{eq:chiral_LO}
\mathcal{L}_{\chi \text{PT}}^{\text{ext}} \supset &-i g_x X_\mu (x_V^u-x_V^s) \left(\partial^\mu K^- K^+ -\partial^\mu K^+ K^- \right)\nonumber\\
    &  -i X_\mu g_x(x_V^u-x_V^d) \left(\partial^\mu \pi^- \pi^+ -\partial^\mu \pi^+ \pi^- \right)\nonumber\\
    & + \left[-i g_x X_\mu x_V^{32}\,\left(\partial^\mu K^+ \pi^--\partial^\mu\pi^-K^+\right) +\text{h.c.}\right] \, .
\end{align}

Some comments are in order to be made about this result:

\begin{itemize}
\item All the couplings in eq. \eqref{eq:chiral_LO} are vectorial in nature. This is a consquence of the fact that the matrix element of the axial-vector component of the quark bilinears in eq. \eqref{eq:X_Lag_with_SM_Fermions} between external pseudo-scalar states is null;
\item In the limit of universal vector couplings, \textit{i.e.} $x_V^u = x_V^d = x_V^s$, both the $K^+ K^- X$ and the $\pi^+ \pi^- X$ interaction terms vanish due to the underlying $SU(3)_V$ chiral symmetry. Contrarily, since flavour-changing currents are not conserved, the $K^\pm \pi^\mp X$ vector coupling does not vanish.
\end{itemize}

It is important then to notice that $\mathcal{O}(p^2)$ contributions to $\Delta S = 1$ processes such as $K^\pm \rightarrow \pi^\pm X$ can be generated only if the vectorial couplings have a non-null off-diagonal entry $x_V^{32}$. 
If this is absent at tree level, it can nonetheless be generated at one-loop level by the exchange of a virtual W boson and an up-type quark \cite{Dror:2017ehi, Dror:2017nsg}.

Weak interactions however do not limit themselves to provide one-loop effects to the decay process we are considering, but they generate as well tree-level effects once higher-order terms in the momentum expansion that are the chiral equivalent of four-fermion operators are taken into account. 

The analysis of such contributions will be the topic covered by the next subsection.

\subsection{Chiral Lagrangian for weak interactions}

In the SM, at energies above the chiral symmetry breaking scale, 
$\Delta S=1$ transitions are induced by the effective 
Lagrangian~\cite{Buchalla:1995vs}
\begin{equation}
\label{eq:SMDelta1Lag}
    \mathcal{L}^{\Delta S= 1}_{\rm SM} = G \sum_{i=1}^{10} C_i(\mu) O_i (\mu) \qquad \text{with } \qquad G \equiv -\frac{G_F}{\sqrt{2}}V_{ud}V^*_{us} \, , 
\end{equation}
where 
%
\begin{equation}
\label{eq:DeltaS1Operators}
    \begin{array}{ll}
        Q_1 = 4(\bar{s}_L\gamma_\mu d_L) (\bar{u}_L \gamma_\mu u_L),  & \qquad \qquad Q_2 = 4(\bar{s}_L\gamma_\mu u_L) (\bar{u}_L \gamma_\mu d_L), \\
        Q_3 = 4(\bar{s}_L\gamma_\mu d_L) (\bar{q}_L \gamma_\mu q_L),  & \qquad \qquad Q_4 = 4(\bar{s}^\alpha_L\gamma_\mu d^\beta_L) (\bar{q}^\beta_L \gamma_\mu q^\alpha_L), \\
        Q_5 = 4(\bar{s}_L\gamma_\mu d_L) \sum_q(\bar{q}_R \gamma_\mu q_R),  & \qquad \qquad Q_6 = 4(\bar{s}^\alpha_L\gamma_\mu d^\beta_L) \sum_q(\bar{q}^\beta_R \gamma_\mu q^\alpha_R), \\
        Q_7 = 6(\bar{s}_L\gamma_\mu d_L) \sum_q e_q(\bar{q}_R \gamma_\mu q_R),  & \qquad \qquad Q_8 = 6(\bar{s}^\alpha_L\gamma_\mu d^\beta_L)\sum_qe_q (\bar{q}^\beta_R \gamma_\mu q^\alpha_R), \\
        Q_9 = 6(\bar{s}_L\gamma_\mu d_L) \sum_q e_q(\bar{q}_L \gamma_\mu q_L),  & \qquad \qquad Q_{10} = 6(\bar{s}^\alpha_L\gamma_\mu d^\beta_L)\sum_q e_q (\bar{q}^\beta_L \gamma_\mu q^\alpha_L),  \\
    \end{array}
\end{equation}
%
$q = {u,d,s}$, $e_u=2/3$ and $e_d = e_s = -1/3 $; $\alpha$ and $\beta$ are colour indices which, if unspecified, are understood to be contracted between the two quarks in the same current.

The construction of the chiral counterpart to  eq. \eqref{eq:DeltaS1Operators} proceeds in two steps: 
\begin{itemize}
\item Firstly, one constructs the chiral structures describing the product of two fermionic currents. These structures must possess the same chiral transformation properties of the corresponding quark currents and are obtained by exploiting the quark-hadron duality between the Lagrangians of eqs.~\eqref{eq:QCD_Lag_With_Sources} and \eqref{eq:Chipt_Lag_With_Sources}, valid at low energies.

One can then find the chiral counterparts to the various Dirac structures by taking appropriate functional derivatives of the QCD and the $\chi$PT actions with respect to the same external sources.

\item 
The product of quark currents can then be decomposed into the irreducible representations of the flavour algebra. This is done by defining appropriate projectors which have to be applied as well to the chiral realisation of the quark currents. In this way one can obtain a set of operators in the chiral theory that are automatically classified according to the irreducible representation of the flavour algebra they belong to and that can be thus directly related to the initial ones, expressed in terms of quark bilinears (see e.g.~\cite{Pich:2021yll, RBC:2001pmy, Lehner:2011fz}). 

\end{itemize}

Once this two-step program is carried out, one can finally reproduce the $\Delta S = 1$ chiral Lagrangian of ref.~\cite{Pich:2021yll}, 
which takes the following simple form 

\begin{equation}
    \label{eq:ChiptWeakSM}
\begin{split}
    \mathcal{L}^{\Delta S= 1}_{\text{eff}}=G f_\pi^4 &\big\{ g_{27} \left(L^3_{\mu,\,2}L^{\mu,\,1}_1+\frac{2}{3}L^1_{\mu,\,2}L^{\mu,\,3}_1-\frac{1}{3}L^3_{\mu,\,2}\text{tr}\,{\left[L^\mu\right]})\right)+  g_8^S \, L^3_{\mu,\,2}\text{tr}\,{\left[L^\mu\right]} \\ 
    &+   g_8\left(\text{tr}\,{\left[\lambda L_\mu L^\mu\right]}+e^2 g_{\text{ew}}f_\pi^2\text{tr}\,\left[\lambda U^\dagger Q U\right]\right)\big\} \, , 
\end{split}
\end{equation}
Here $\lambda \equiv \frac{1}{2} (\lambda_6-i\lambda_7)$ is responsible for the $s\rightarrow d$ flavour transition and we have specialised $Q=\frac{1}{3} \text{diag}(2, -1, -1)$ to be the charge matrix for quarks. The left-handed current chiral $L_\mu$ is defined via $ L_\mu \equiv i U^\dagger D_\mu U$.
Out of the pieces making up eq.~\eqref{eq:ChiptWeakSM}, the first one transforms in the $(27_L, 1_R)$ representation of the flavour group, while the second and the third ones transform in the $(8_L, 1_R)$ and $(8_L, 8_R)$ representation, respectively.
Clearly, no singlet term can have any effect on $\Delta S = 1$ transitions.
The $\mathcal{O}(1)$ coefficients $g_{27}$, $g_8$, $g_8^S$ and $g_{\text{ew}}$ are functions of non-perturbative effective parameters, as well as of the Wilson coefficients of the weak operators, see eq.~\eqref{eq:SMDelta1Lag}. 
Expanding ~\eqref{eq:ChiptWeakSM} and keeping only the contributions relevant for our analysis, we find 
 
\begin{equation}
\begin{split}
    \label{eq:ChiPT27}
    \mathcal{L}^{\Delta S= 1}_{\text{eff}} 
    \supset & \,\,\frac{2}{3}f^2 g_{27} G \left(2\partial^\mu K^+ \partial_\mu \pi^- + g_x X_\mu\left[i\partial^\mu K^+ \pi^- (4x_A^u-x_A^d-3x_A^s+2x_V^u-2x_V^d)\right.\right.  \\ 
    &\left. \left. -i\partial^\mu\pi^- K^+ (4x_A^u-3x_A^d- x_A^s+2x_V^u -2x_V^d)+\text{h.c.}\right] \right)\\
    &+ 2f^2 g_8^S G g_x \, (x_A^u+x_A^d+x_A^s)\, X_\mu  \left[i \,\left(\partial^\mu K^+ \pi^--\partial^\mu\pi^-K^+\right) +\text{h.c.}\right] \\
    &+ 2f^2 g_8 G \left(\partial^\mu K^+ \partial_\mu \pi^- + g_x X_\mu\left[i\partial^\mu K^+ \pi^- (x_A^u+x_A^s+x_V^u-x_V^d)\right.\right.  \\
    &\left. \left. -i\partial^\mu\pi^- K^+ (x_A^u+x_A^d+x_V^u-x_V^s)+\text{h.c.}\right] \right) +
    2 f^4 G e^2 g_8 g_{\text{ew}} K^+ \pi^- \, , 
\end{split}
\end{equation}
which includes both a $K\pi$ mixing term and a flavour-violating $K^\pm\to\pi^\pm X$ interaction.  

Interestingly enough, one is now sensitive to both vectorial and axial couplings since the hadronic matrix element $\bra{K}O_i \ket{\pi}$ -with $O_i$ from eq. \eqref{eq:DeltaS1Operators}- receives contributions from both vector and axial-vector currents.

\section{$K^\pm \rightarrow \pi^\pm X$ in $\chi$PT}

The lagrangian pieces in eqs. \eqref{eq:chiral_LO} and \eqref{eq:ChiptWeakSM} can be used in order to compute the decay rate for the process $K^\pm \rightarrow \pi^\pm X$. 

\begin{figure}[ht]
\centering
\begin{tikzpicture}[]
\begin{feynman}[]
\vertex(a);
\node[crossed dot,right = 1.5 cm of a](b);
\vertex[below left= 0.7 cm of b](n);
\vertex[right = 1.7 cm  of b](c);
\vertex[below right = of b](d);
\diagram*{
(a) -- [charged scalar, edge label=$K^+$] (b),
(d) -- [boson, edge label=$X$] (b) ,
(b) -- [charged scalar, edge label=$\pi^+$] (c),
};
\end{feynman}
\end{tikzpicture}\qquad
\begin{tikzpicture}[]
\begin{feynman}[]
\vertex(a);
\node[crossed dot,right = 1 cm of a](b);
\vertex[below left= 0.7 cm of b](n);
\node[dot, right = 1 cm of b](c);
\vertex[right = 1 cm  of c](e);
\vertex[below right = of c](d);
\diagram*{
(a) -- [charged scalar, edge label=$K^+$] (b),
(b) -- [charged scalar, edge label=$\pi^+$] (c),
(d) -- [boson, edge label=$X$] (c) ,
(c) -- [charged scalar, edge label=$\pi^+$] (e),
};
\end{feynman}
\end{tikzpicture}\qquad
\begin{tikzpicture}[]
\begin{feynman}[]
\vertex(a);
\node[crossed dot,right = 1 cm of b](c);
\vertex[below right= 0.7 cm of c](n);
\node[dot, right = 1 cm of a](b);
\vertex[right = 1 cm of c](e);
\vertex[below right = of b](d);
\diagram*{
(a) -- [charged scalar, edge label=$K^+$] (b),
(b) -- [charged scalar, edge label=$K^+$] (c),
(d) -- [boson, edge label=$X$] (b) ,
(c) -- [charged scalar, edge label=$\pi^+$] (e),
};
\end{feynman}
\end{tikzpicture}
\caption{Diagrams generating the tree-level transition $K^\pm\to\pi^\pm X$ in $\chi$PT, see ref. \cite{DiLuzio:2023xyi}.}
\label{fig:FeynmanDiagrams}
\end{figure}
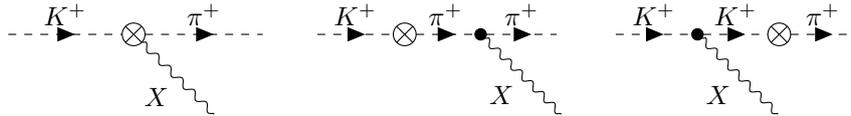

The Feynman diagrams describing the process under consideration are depicted in fig. \ref{fig:FeynmanDiagrams}: the $X$ boson can be either emitted at the same vertex where the flavour transition takes place (first diagram) or at a different one (second and third diagrams). In the second case, weak interactions prompt a flavour transition while the $X$ boson is radiated at a different interaction point from an external leg.

A pretty simple expression for the decay rate can be found assuming generation universality of the couplings ($x_{V,A}^u = x_{V,A}^d = x_{V,A}^s$) and taking the limit $m_K \gg m_X, m_{\pi}$
\begin{align}
\label{eq:gamma_simplified}
\begin{split}
    \Gamma \approx \frac{m_K}{2\pi} \left(\frac{m_K}{m_X}\right)^2 G_F^2 f_{\pi}^4 \, |V_{us}|^2 \, g_x^2 \, (x_A^u)^2 
    \left(g_8 + \frac{3}{4} g^S_8 \right)^2\, .
\end{split}
\end{align}

It is interesting to notice that as a consequence of the $SU(3)_V$ chiral symmetry, in the limit of universal vector couplings, the decay rate of $K^\pm\to\pi^\pm X$ becomes independent of these couplings.
Secondly, it has to be appreciated that the expected enhancement factor $(m_K/m_X)^2$ in eq.~\eqref{eq:gamma_simplified} for small $m_X$ is correctly recovered, and is here produced by the longitudinal component of the polarization vector: $\sum\varepsilon_\mu^*(q)\varepsilon_\nu(q)=-\eta_{\mu\nu}+\frac{q_\mu q_\nu}{m_X^2}$.

The one-loop effects discussed in \cite{Dror:2017nsg} can be incorporated in eq. \eqref{eq:X_Lag_with_SM_Fermions} via 
\begin{equation}
\label{eq:shift}
x^{32}_V \to x^{32}_V - x^{\rm{eff}}_{sd}\, 
\end{equation}
where, in the limit of universal couplings, \textit{i.e.} $x_{V,A}^{u_i} = x_{V,A}^d = x_{V,A}^s$, and keeping only the dominant loop effects stemming from the exchange of the top quark, we obtain
\begin{align}
\label{eq:OneLoopFlavourUniversal}
x^{\rm{eff}}_{sd} &\simeq 
\frac{g^2}{64\pi^2}\,V_{td}V^*_{ts} 
x_A^u f(x_t) 
\end{align}
with
\begin{align}
f(x_t) = x_t
\left[\frac{2}{\epsilon}+\log{\frac{\mu^2}{m_t^2}}-\frac{1}{2}-3
\frac{(1-x_t+\log{x_t})}{(x_t-1)^2}\right] \, .
\end{align}

This allows us to compare tree-level vs loop-induced effects, by studying the ratio
\begin{align}
\label{eq:1_loop_vs_tree}
\frac{x^{\rm{eff}}_{sd}}{4 g_8 f_{\pi}^2 G x_A^u} \approx f(x_t) \, ,
\end{align}
where $f(x_t)$ is a model-dependent loop function which depends on the specific UV completion of the effective theory and that is expected to be of order $\mathcal{O}(1)$.

Loop- and tree- level effects are thus seen to be comparable in magnitude. However, the former depend critically on the specifics of the UV completion of the theory, whereas the latter provide robust and model-independent results.

\subsection{Flavour bounds vs.~beam-dump and collider searches}
\label{sec:numerics}

 \begin{figure}[t]
        \centering
        \includegraphics[width=0.7\textwidth]{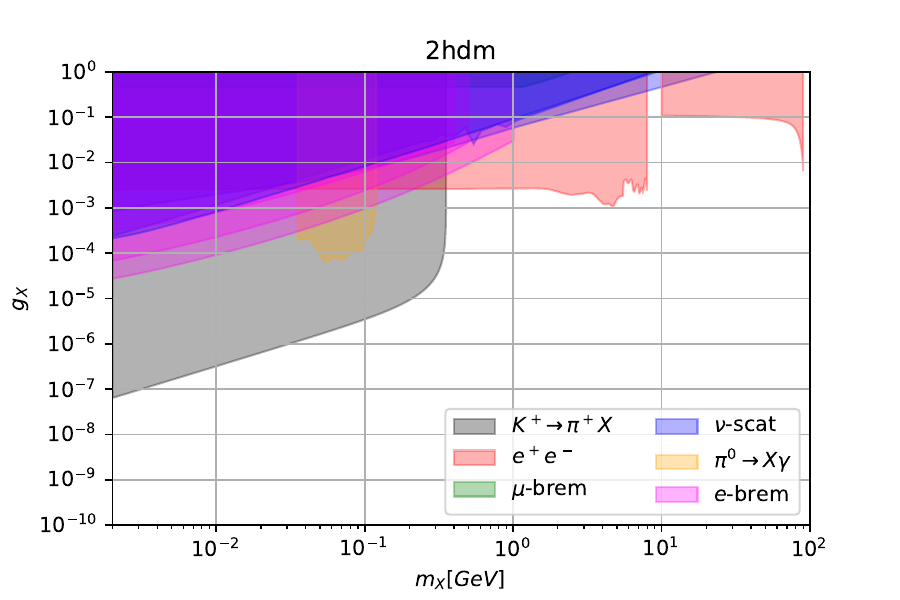} \\
        \caption{The dark shaded area represents the 
        tree-level $K^\pm\to\pi^\pm X$ bound obtained in ref. \cite{DiLuzio:2023xyi}. 
        Limits from beam-dump and collider searches are 
        obtained with DarkCast \cite{Baruch:2022esd} 
        and are shown for the purpose of comparison 
        for the three benchmark models 
        given in Table \ref{tab:model charges}.
        }
        \label{fig:Bounds}
    \end{figure}

    The results of the previous section can be employed in order to explore the capability of the process $K^\pm\to\pi^\pm X$ 
    to probe new light vector bosons.
    The DarkCast package ~\cite{Ilten:2018crw, Baruch:2022esd} enables one to derive bounds on vector and axial couplings of NP scenarios featuring new spin-1 particles by imposing current and future experimental constraints on several processes.
    The bounds in the $(m_X,g_x)$ plane arising from a variety of beam-dump and collider searches~\cite{Baruch:2022esd} as well as from the flavour changing process $K^\pm\to\pi^\pm X$ discussed in this paper are shown in fig.~\ref{fig:Bounds}.  
    The plot refers to the benchmark two-Higgs doublet model in~\cite{Baruch:2022esd}, with charge assignment $x_V^e = 0.044$, $x_V^\nu = 0.05$, $x_V^{u,c,t} = 1.021$, $x_V^{d,s,b} = 0.015$, $x_A^e = -0.1$, $x_A^\nu = 0.05$, $x_A^{u,c,t} = -0.95$ and $x_A^{d,s,b} = -0.1$. 
%

    The bounds from the process $K^\pm\to\pi^\pm X$ are obtained by assuming tree-level, flavour-diagonal (\textit{i.e.} disregarding the one-loop effects) couplings in eq.\eqref{eq:X_Lag_with_SM_Fermions}, and exploiting the measurement of $\text{BR}(K^+\to \pi^+\nu\nu)=( 1.73^{+1.15}_{-1.05})\times 10^{-10}$ by the E949 experiment at BNL~\cite{E949:2008btt}.
    In particular, we imposed the $2\sigma$ bound $\text{BR}(K^+\to \pi^+X)\lesssim 4 \times 10^{-10}$.
    Remarkably, in all scenarios of fig.~\ref{fig:Bounds}, the process $K^\pm\to\pi^\pm X$ 
    sets the strongest to date model-independent bound in the $(m_X,g_x)$ plane for $m_X < m_K - m_\pi$.

\section{Conclusions}

Among the most studied scenarios for new physics beyond the Standard Model are the ones introducing a new, feebly interacting massive particle. 
A particularly interesting subclass of these models features light spin-1 bosons having masses smaller than a few GeVs. Such a possibility has been extensively analysed in the light of experimental searches at colliders and at beam-dump experiments.
However, considerably less attention has been given to the flavour constraints by the rare decay $K^\pm \rightarrow \pi^\pm X$, which is the object of our work.

We extended previous analyses by building the most general $\Delta S = 1$ chiral Lagrangian as induced by the SM weak interactions up to order $\mathcal{O}(p^4)$. 
In particular we observe that the $\mathcal{O}(p^2)$ terms in such a Lagrangian describe the loop-induced effects from \cite{Dror:2017nsg} to the decay process under consideration, while the $\mathcal{O}(p^4)$ terms generate the flavour transition already at the tree-level.
Due to a different $\lambda$ suppression ($\lambda$ is here the Wolfenstein parameter), the two effects turn out being comparable in strength. 
However, whereas the loop-induced effects suffer from a dependence from the details of the explicit UV mechanism providing the spin-1 boson with a mass, the tree-level ones are model-independent.

With our work we showed that the flavour process $K^\pm \rightarrow \pi^\pm X$ puts the strongest model-independent constraints on the diagonal axial-vector couplings to light quarks with a NP light spin-1 particle $X$ in the mass range $m_X < m_K-m_\pi$.

\begin{small}

\bibliographystyle{varenna}
\bibliography{sample.bib}

\end{small}

\end{document}